\documentclass[12pt]{article}
\usepackage{a4}
\usepackage{epsfig}

\begin{document}
\begin{titlepage}
\begin{flushright}
LU TP 99-05\\
February 1999
\end{flushright}
\vfill
\begin{center}
{\Large\bf A Parametrization for $K^+\to \pi^+\pi^- e^+\nu$.$^\dagger$}
\\[2cm]
{\bf Gabriel Amor\'os$^{a,b}$ and Johan Bijnens$^a$}\\[1cm]
$^a$ Department of Theoretical Physics 2, Lund University\\
S\"olvegatan 14A, S 22362 Lund, Sweden\\[1cm]
$^b$ Department of Physics, P.O. Box 9\\
FIN-00014 University of Helsinki, Finland
\end{center}
\vfill
\begin{abstract}
We discuss various models and Chiral Perturbation Theory results
for the $K_{l4}$ form factors $F$ and $G$. We check in how much
a simple parametrization with a few parameters can be used to
extract information from experiment.
\end{abstract}

\vfill
\footnoterule
{$^\dagger$\small Work
supported in part by TMR, EC-Contract No. ERBFMRX-CT980169
(EURO\-DA$\Phi$NE).}
\end{titlepage}
\section{Introduction}

In this note we give a simple parametrization and behaviour with few 
kinematical variables of the form factors contributing to the decay 
$K^+\to \pi^+\pi^- e^+\nu$.   

The interest of this work resides in the new measurements 
of $K\to\pi\pi\ell\nu$ ($K_{l4}$) to be obtained with 
new facilities in the inmediate future. The increase in the
number of events obtained 
will give more information about the dependence with the kinematical 
variables, generalizing and improving the previous and old results 
\cite{rosselet}.
This decay will be a major test of the 
consistency of the Chiral Perturbation Theory (CHPT), comparing the prediction 
of some $O(p^4)$ constants of the chiral expansion with the values given 
by other processes. Even more, this decay is expected to be the best one 
for a precise determination of the isoscalar $\pi \pi$ S-wave phase 
shift near threshold which will allow comparing CHPT and other
approaches to chiral symmetry breaking. A review of the CHPT results
can be found in \cite{kl4daphne} and references therein.

One of the main results from the previous experiment is the observation 
of dependence on the energy of the dipion system ($s_\pi$).
It was consistent with a linear dependence
for all the form 
factors and with the same slope. The contribution from other 
kinematical variables was not considered. However, as we will see later, 
the expected variations of the form factors on the energy 
of the leptonic pair ($s_\ell$) and on the 
$\cos \theta_\pi$, is on the same level as the errors in the previous
experiment and will probably become important for the next generation
of experiments.
In \cite{rosselet} the form factors were fitted with a linear dependence
only and the same slope for all form factors. The purpose of this note
is to examine various theoretical expectations and check how well
a simple extension of this parametrization performs. This will allow
then an improved extraction of the $\pi\pi$ phase shifts from the data.

Papers using CHPT and extensions for $K_{l4}$ are \cite{BCG} and
\cite{chptp4}. A review of early work is \cite{Chounet}. An alternative
method to extract $\pi\pi$ phases is described in \cite{Pais,PaisTreiman}
and references therein but it does not allow to use the full experimental
information as much as our parametrization allows.

The paper is divided as follows: we introduce the form factors
in Sect. \ref{defform}. We then discuss the singularities that appear
in the various cuts in Sect. \ref{cuts}. In Sect. \ref{resonances}
we study two different models for inclusion of resonances, in Sect. 
\ref{CHPTp4} CHPT at one loop and with inclusion of part of
the $p^6$ effects in Sect. \ref{doublelog}. We then discuss also some
unitarization effects and finally the $H$ form factor as well.
Sect. \ref{parametrization} contains our proposal for the parametrization
of the form factors.
The last section contains our conclusions.

\section{Form Factors}
\label{defform}

The decay $K^+(p)\to \pi^+(p_1)\pi^-(p_2) e^+(p_\ell)\nu(p_\nu)$ is given 
by the amplitude
\begin{equation}
      T = \frac{G_F}{\sqrt{2}} V^\star_{us} \bar{u} (p_\nu) \gamma_\mu
      (1-\gamma_5) v(p_\ell) (V^\mu - A^\mu)
      \label{k11}
\end{equation}
where $V^\mu$ and $A^\mu$ are parametrized in terms of four form factors:

\begin{eqnarray}
V_\mu & = & - \frac{H}{M^3_K} \epsilon_{\mu \nu \rho \sigma} L^\nu
      P^\rho Q^\sigma \;,
      \nonumber \\
      A_\mu & = & -i\frac{1}{M_K} \left [ P_\mu F +
      Q_\mu G + L_\mu R \right ]
\end{eqnarray}
with
\begin{equation}
 P = p_1 + p_2 \;, \;  Q = p_1 - p_2 \;, \; L = p_\ell + p_\nu \;, 
\; N = p_\ell - p_\nu \;.
\end{equation}
and $\epsilon_{0123}=1$. $\bar{u}$ and $v$ are the lepton Dirac spinors and
$G_F$ is the Fermi constant.

In this work we consider the theoretical predictions for the 
dependence of these form factors $F$, $G$ and $H$, 
with the kinematical variables 
$s_\pi=(p_1+p_2)^2$, $s_\ell=(p_\ell+p_\nu)^2$ and $\cos \theta_\pi$ ($\theta_\pi$ 
is the angle of the $\pi^+$ in the $\pi\pi$ system respect to the dipion line 
of flight). The $R$-form factor is negligible in decays with
an electron in the final state and we will not discuss it.

The relation between form factors and the width can be found in 
several references \cite{kl4daphne,BCG,Pais}.

The main objective of this paper is to check in the existing models and
calculations in how much simple parametrizations of $F$, $G$ and $H$
are sufficient. We will basically check if linear dependence
on $s_\pi$, $s_\ell$ and $\cos\theta_\pi$ is sufficient or not.
\section{Cut Structure}
\label{cuts}

\begin{figure}
\begin{center}
\epsfig{file=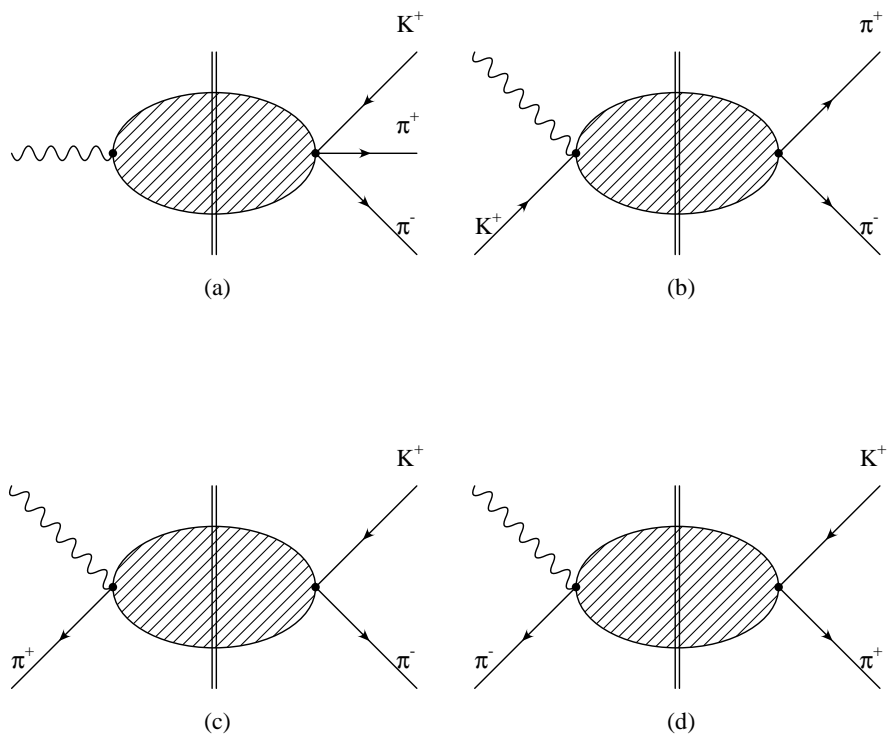,width=12cm}
\end{center}
\caption{\label{figcuts}The possible cuts in the decay $K^+\to\pi^+\pi^-
e^+\nu$. The double line is the cut. The wiggly line is the weak current.}
\end{figure}

One indication that a constant or a linear structure is enough to describe a
given form factor is, how far the masses of the possible
intermediate states are from the physical
domain. 
This is done by looking at the states contributing to the various cuts.
The possible cuts with strong interaction intermediate states are
shown in Fig. \ref{figcuts}.
We discuss now the cuts in order of the figure.

\subsection{Cut (a)}

The strong interaction intermediate states that can couple here
are in order of their mass:
$K$, $K\pi$, $K^*$, $K\pi\pi$, $K_0$, $K_1$,\dots.

The $K$ and $K_0$ intermediate states can only contribute to the
form factor $R$ which is negligible in this decay. The $K\pi$ and $K^*$
intermediate state are vector states and only contribute to the form factor
$H$. Their effect has been estimated for $K\pi$ by one-loop CHPT
and for the $K^*$ by a VMD estimate using the hidden gauge vector model
in \cite{Amettler}. There was in effect a large cancellation between
the two effects leading to a rather constant form factor across the relevant
phasespace. A linear aproximation for both 
amplitudes separately should in any case be
sufficient because the relevant momentum, $s_\ell$ is far 
from the thresholds.

The $K\pi\pi$ has a threshold of about 0.765~GeV which is rather far
from the physical region which ends at about 0.225~GeV. In addition in
CHPT this only starts at the $p^6$ level.
There are two $K_1$ resonances, one at 1.27~GeV and one at 1.4~GeV. Both are
again rather far from the physical region so we expect that their effect will
also be small and certainly describable by at most a linear function of $s_\ell$.

If one checks the contribution of the axials of \cite{EGPR} we also
notice that the axial-vectors only start contributing to this process
at order $p^6$.

In conclusion, we expect the effects of this cut for $K_{e4}$ decays
to be linear and fairly small.

\subsection{Cut (b)}

Here the possible intermediate states are in order of mass again:
$\pi\pi$, $\rho$, $f_0$,\ldots. The $\pi\pi$ intermediate
state is in fact one of the major reasons to study this decay in
order to learn more about $\pi\pi$ phase shifts. The cut here is
in the physical region for this decay. The effects of $\pi\pi$ final state
rescattering will be discussed later, see also \cite{kl4daphne,BCG}.

The effects of the $\rho$ are confined mainly to the form factor
$G$. There is a very small curvature in $G$ due to this intermediate state
visible in estimates of its effect, although in the region with
sufficient data it is rather small. See section \ref{resonances} for its 
effect.

\subsection{Cut (c)}

This channel has charge 2 and has thus no resonance enhancement.
The relevant intermediate state is mainly $K\pi$ and this also
has a small phase shift in the relevant channel. We thus expect it
to be well described by oneloop CHPT.

\subsection{Cut (d)}

Possible intermediate states here are:
$K\pi$, $K^*$, $K_0$,\ldots.
The relevant variable in this channel is $t$ and it is fairly far away from
the threshold. Its effects are thus expected to be well described
by a linear function in $t$, or linear in $s_\ell$ and $\cos\theta_\pi$.
Nonlinear effects on $s_\ell$ are at most of order $s_\ell^2/m_{K\pi}^4$
which is about 1\%. Actual estimates from CHPT at one-loop
and various models confirm this expectation.

\subsection{Other singularities}

In addition to the above real singularities there are also kinematical
singularities present in the various quantities. An example of this is
the singularities present in the partial waves for $\mbox{Re}\,s_\pi\le 0$.
These follow because the values of $t$ can become larger than $m_K+m_\pi$
for some values of $\cos\theta_\pi$ and $s_\ell$.

\section{Resonances}
\label{resonances}

\subsection{Resonances in a vector--axial-vector dominance model}

In this section we use the model of Ref. \cite{FM} for $\tau$ decays.
Basically this model takes vector and axial-vector mesons and writes
the amplitude such that in the low-energy limit all constraints
from chiral symmetry are satisfied. In all channels contributions
from the two lowest channels are included and the meson propagators
are described by Breit-Wigner functions including a $s$-dependent width.

\begin{figure}
\begin{center}
\epsfig{file=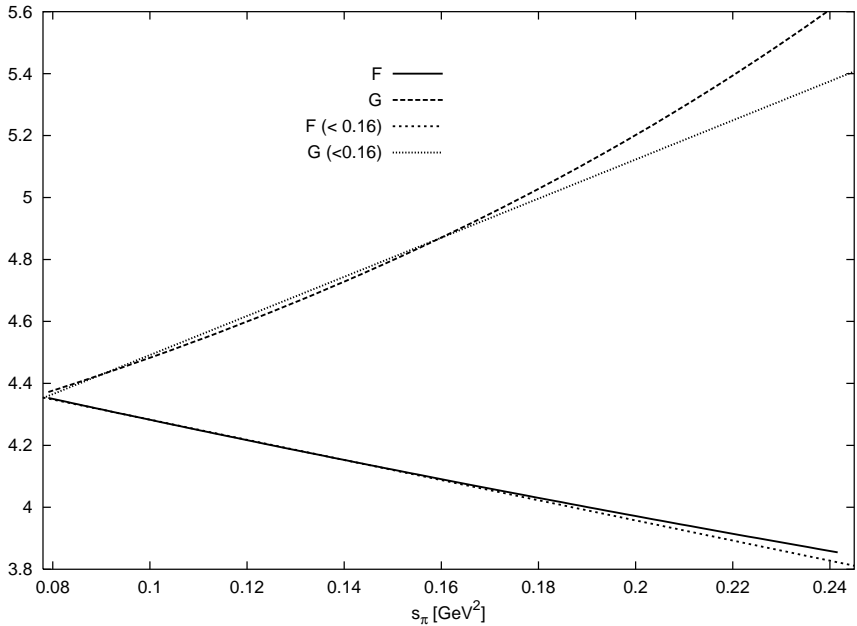,width=12cm}
\end{center}
\caption{\label{figres1}
The form factor $F$ and $G$ of the model of Ref. \cite{FM}
as a function of $s_\pi$ at $\cos\theta_\pi=s_\ell=0$.}
\end{figure}
In Fig. \ref{figres1} we have shown the $F$ and $G$ form factors of this
model as a function of $s_\pi$ for $\cos\theta_\pi=s_\ell=0$.
In accordance with the discussion in Sect. \ref{cuts} we see that
there is very little curvature. The curvature visible in $G$ is due to
the $\rho$-pole in this model.
We have checked this by setting $m_\rho=1.37~$GeV in the model and then all
curvature disappears.
Experimentally there are very few events\footnote{This will also be true
in the next generation of experiments. We have in Fig. \ref{figbcg}
shown the data of \cite{rosselet} with the points plotted at
the average energy in the bins used.}
above $s_\pi=0.16~$GeV$^2$ or $E_{\pi\pi}\le0.4~$GeV. In the figure the
fits linear in $s_\pi$ to the model curves below this energy are also shown.
As can be seen the linear approximation is in the relevant region
quite sufficient as an approximation to this model.

Let us now turn to the $s_\ell$ dependence within this model.
\begin{figure}
\begin{center}
\epsfig{file=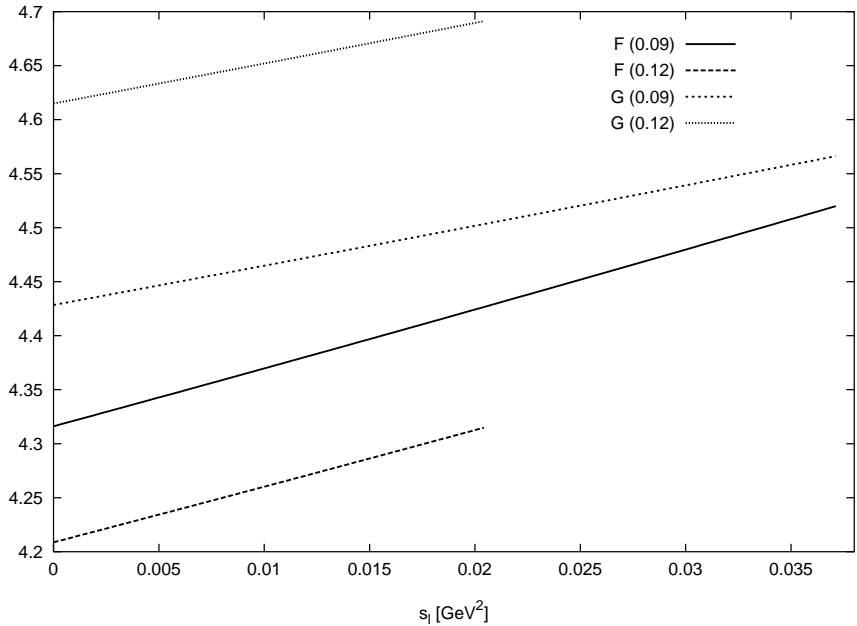,width=12cm}
\end{center}
\caption{\label{figres2}
The form factor $F$ and $G$ of the model of Ref. \cite{FM}
as a function of $s_\ell$ at $\cos\theta_\pi=0$ and
for $\sqrt{s_\pi}= 0.3~$GeV and $\sqrt{s_\pi}= 0.35~$GeV.}
\end{figure}
In Fig. \ref{figres2} the form factors $F$ and $G$ are plotted
for two values of $s_\pi$.
These values are in the region where accurate data can be expected.
The lines are only plotted for the possible
kinematical domains accessible in $K_{e4}$. As is obvious from the
figure the $s_\ell$ dependence in this model is very linear and
any curvature can be neglected
within the expected experimental accuracy.
The differences in the slope for the two values of $s_\pi$ is
small.

We now turn to the $\cos\theta_\pi$ dependence within this model.
\begin{figure}
\begin{center}
\epsfig{file=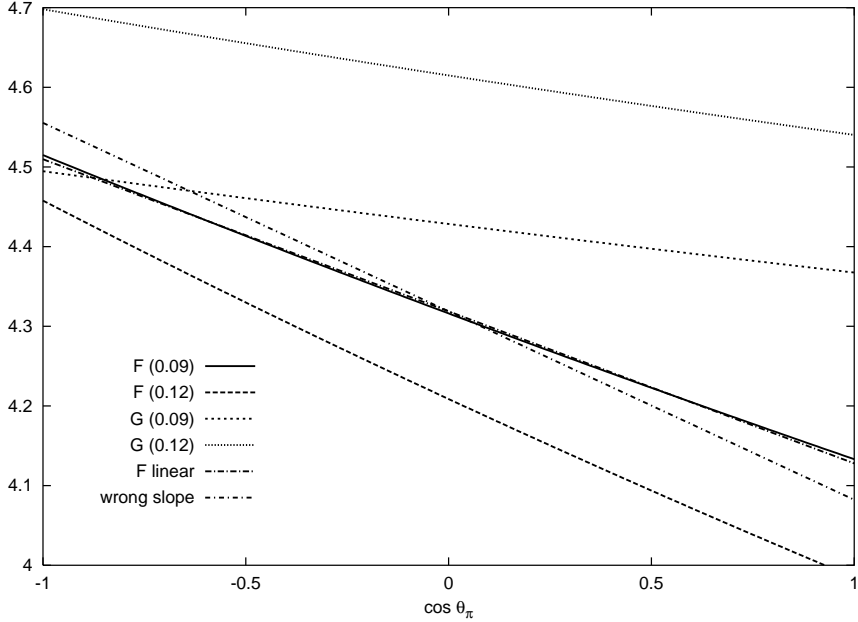,width=12cm}
\end{center}
\caption{\label{figres3}
The form factor $F$ and $G$ of the model of Ref. \cite{FM}
as a function of  $\cos\theta_\pi$ at $s_\ell=0$ and
for $\sqrt{s_\pi}= 0.3~$GeV and $\sqrt{s_\pi}= 0.35~$GeV.}
\end{figure}
In Fig. \ref{figres3} the form factors $F$ and $G$ are plotted
for two values of $s_\pi$. For $G$ it is extremely linear while for $F$
a very small curvature is present. The latter is however below
the expected experimental uncertainties and show that the effect of
$D$-waves\footnote{Meant is here and below, $D$ waves for $F$ and
higher than $D$ waves for $G$.}
is very small in this model. A linear best fit to $F$ is also
shown. The slope changes somewhat between the two values of $\sqrt{s_\pi}$.
The linear best fits are :
\begin{eqnarray}
\label{fpFM}
F(\sqrt{s_\pi}=0.3~\mbox{GeV}) &=& 4.319-0.191 \cos\theta_\pi\nonumber\\
F(\sqrt{s_\pi}=0.35~\mbox{GeV}) &=& 4.213-0.236 \cos\theta_\pi\nonumber\\
G(\sqrt{s_\pi}=0.3~\mbox{GeV}) &=& 4.429-0.064 \cos\theta_\pi\nonumber\\
G(\sqrt{s_\pi}=0.35~\mbox{GeV}) &=& 4.616-0.079 \cos\theta_\pi
\end{eqnarray}
The differences in the $\cos\theta_\pi$ part
at the different energies are visible
but do probably remain within the errors of the next generation of
experiments. This can be seen in Fig. \ref{figres3} where we have for $F$
also plotted the linear fit with the slope of the curve at
$\sqrt{s_\pi}=0.35~$GeV with the central value of the curve at
$\sqrt{s_\pi}=0.3~$GeV. See also below for an approximate description of
the $s_\pi$ dependence.

So for this model a fit with linear $s_\pi$, a linear $\cos\theta_\pi$
and a linear $s_\ell$
dependence seems sufficient in the region where there will be accurate
data. 

\subsection{Resonances using the model of Ref. \cite{BCG}}
\label{BCGreso}

Here we use the resonance parametrization as used in Ref. \cite{BCG}.
The difference with the previous section are that the resonances here only
contribute at next-to-leading order in CHPT. We have therefore added
the lowest order contribution $m_K/(\sqrt(2)F_\pi)$ to $F$ and $G$ in
order to make the plots more comparable to the previous section.
In this parametrization there is also no contribution from the axial-vectors
to $K_{e4}$, differences due to SU(3) breaking in the meson masses and
couplings are also neglected but here scalars are included as well.
In most cases however the main contribution comes from the vectors.

The conclusions are also the same as in the previous section even if there
are some differences in the numerical estimates of the various coefficients
involved.

\begin{figure}
\begin{center}
\epsfig{file=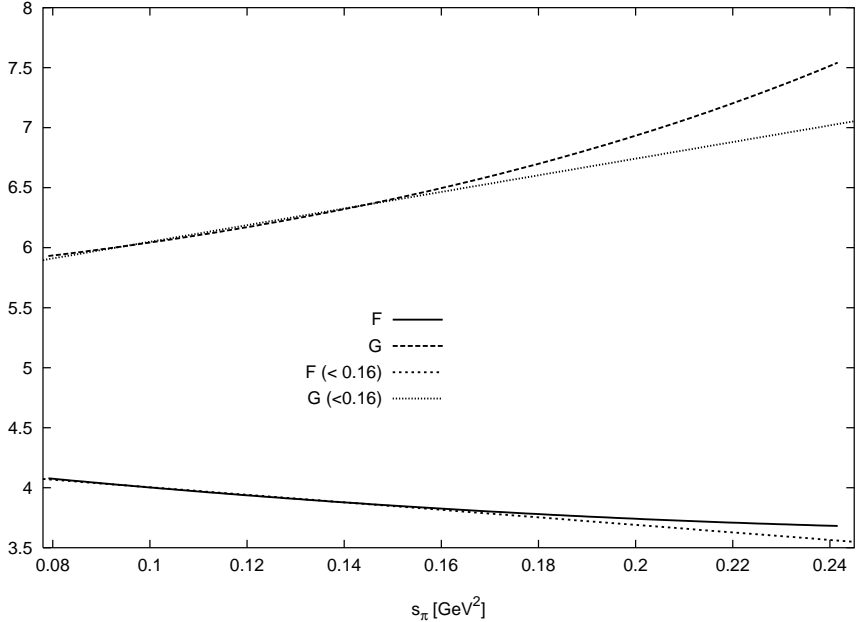,width=12cm}
\end{center}
\caption{\label{figres21}
The form factor $F$ and $G$ of the model of Ref. \cite{BCG}
as a function of $s_\pi$ at $\cos\theta_\pi=s_\ell=0$.}
\end{figure}
In Fig. \ref{figres21} we have shown the $F$ and $G$ form factors of this
model as a function of $s_\pi$ for $\cos\theta_\pi=s_\ell=0$.
In accordance with the discussion in Sect. \ref{cuts} we see that
there is again little curvature. The curvature visible in $G$ is again due to
the $\rho$-pole.
We have checked this by setting $m_\rho=1.37~$GeV in the model and then all
curvature disappears.
Experimentally there are very few events
above $s_\pi=0.16~$GeV$^2$ or $E_{\pi\pi}\le0.4~$GeV. In the figure the
fits linear in $s_\pi$ to the model curves below this energy are also shown.
As can be seen the linear approximation is in the relevant region
quite sufficient as an approximation to this model.

%The $s_\ell$ dependence within this model.
\begin{figure}
\begin{center}
\epsfig{file=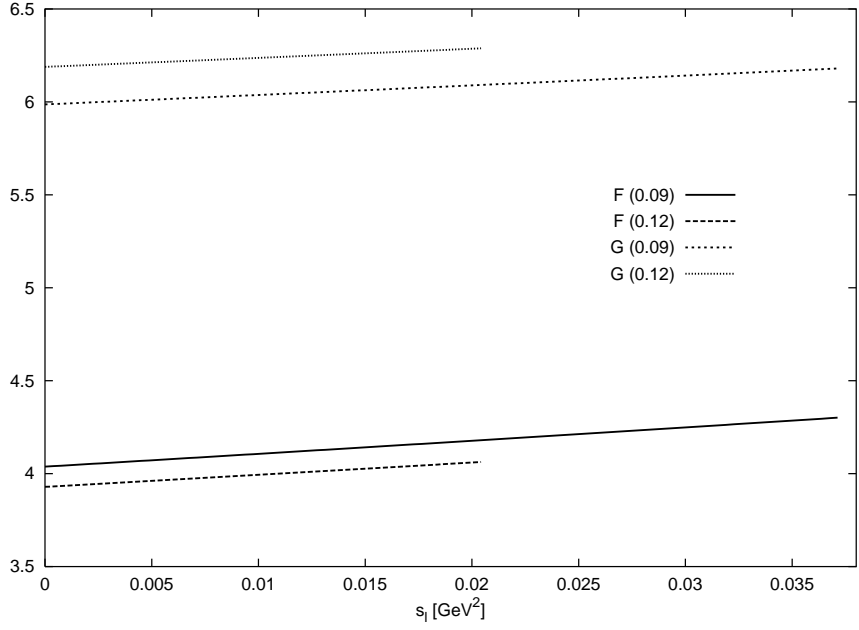,width=12cm}
\end{center}
\caption{\label{figres22}
The form factor $F$ and $G$ of the model of Ref. \cite{BCG}
as a function of $s_\ell$ at $\cos\theta_\pi=0$ and
for $\sqrt{s_\pi}= 0.3~$GeV and $\sqrt{s_\pi}= 0.35~$GeV.}
\end{figure}
In Fig. \ref{figres22} we plot the $s_\ell$ dependence of the 
form factors $F$ and $G$ for two values of $s_\pi$.
The lines are only plotted for the possible
kinematical domains accessible in $K_{e4}$.
The $s_\ell$ dependence is very linear and
any curvature can be neglected
within the expected experimental accuracy.
The differences in the slope for the two values of $s_\pi$ is
small.

We now turn to the $\cos\theta_\pi$ dependence within this model.
\begin{figure}
\begin{center}
\epsfig{file=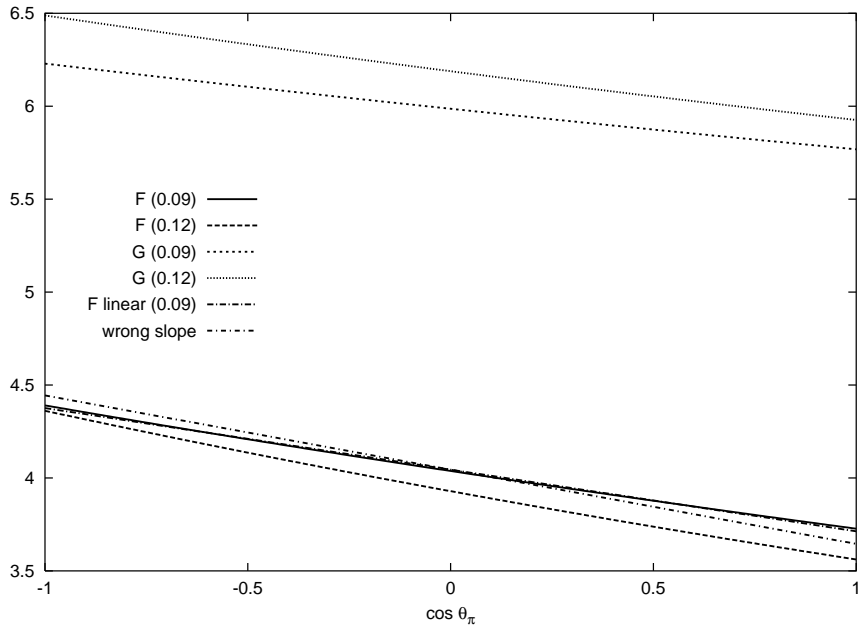,width=12cm}
\end{center}
\caption{\label{figres23}
The form factor $F$ and $G$ of the model of Ref. \cite{BCG}
as a function of  $\cos\theta_\pi$ at $s_\ell=0$ and
for $\sqrt{s_\pi}= 0.3~$GeV and $\sqrt{s_\pi}= 0.35~$GeV.}
\end{figure}
In Fig. \ref{figres23} the form factors $F$ and $G$ are plotted
for two values of $s_\pi$. For $G$ it is extremely linear while for $F$
a very small curvature is present. The latter is however below
the expected experimental uncertainties and show that the effect of
$D$-waves is very small in this model. A linear best fit to $F$ is also
shown. The slope changes somewhat between the two values of $\sqrt{s_\pi}$.
The linear best fits are :
\begin{eqnarray}
\label{fpBCG}
F(\sqrt{s_\pi}=0.3~\mbox{GeV}) &=& 4.045-0.331 \cos\theta_\pi\nonumber\\
F(\sqrt{s_\pi}=0.35~\mbox{GeV}) &=& 3.940-0.399 \cos\theta_\pi\nonumber\\
G(\sqrt{s_\pi}=0.3~\mbox{GeV}) &=& 5.991-0.230 \cos\theta_\pi\nonumber\\
G(\sqrt{s_\pi}=0.35~\mbox{GeV}) &=& 6.195-0.281 \cos\theta_\pi
\end{eqnarray}
The differences in the $\cos\theta_\pi$ part at the different
energies are visible but do probably remain within the errors of the next
generation of experiments.

So for this model again a fit with linear $s_\pi$, a linear $\cos\theta_\pi$
and a linear $s_\ell$
dependence seeme to be sufficient in the region where there will be accurate
data.

\section{One-loop CHPT}
\label{CHPTp4}

In Sect. \ref{resonances} we discussed in two models the effects of the
poles present in the various cuts. Now we switch to a discussion
of the effects of the continuum. In this section we will use
CHPT to order $p^4$. The contributions from the $p^4$ constants
$L_i$ to $F$ and $G$ are linear in the kinematical variables $s_\pi$, $t$
and $u$ thus all curvature present is from the loop diagrams.
To one-loop these diagrams have possible two-particle cuts, the main
ones are the $\pi\pi$ and the $K\pi$ ones as described in Sect. \ref{cuts}.
Other cuts present in the loops are $KK$ and $K\eta$. Their effect
is however very linear in the kinematical variables.
We use here the oneloop calculation of \cite{chptp4} in the notation of
\cite{BCG}. For definiteness we use the values of the $L_i^r$ as
given in the DA$\Phi$NE report\cite{introdaphne} and all plots
plot the real part only. Notice that this corresponds to the unitarized
fit of Ref. \cite{BCG} so the fact that the plots do not agree well
with the data of \cite{rosselet} is due to the unitarization effects.

\begin{figure}
\begin{center}
\epsfig{file=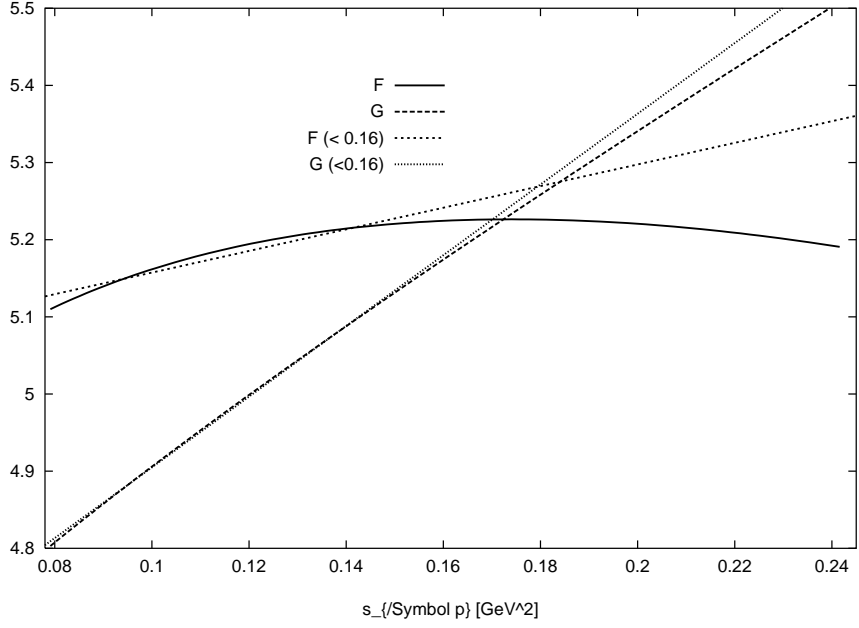,width=12cm}
\end{center}
\caption{\label{figchpt1}
The form factor $F$ and $G$ in $p^4$ CHPT
as a function of $s_\pi$ at $\cos\theta_\pi=s_\ell=0$.}
\end{figure}
In Fig. \ref{figchpt1} we have shown the $F$ and $G$ form factors of this
model as a function of $s_\pi$ for $\cos\theta_\pi=s_\ell=0$.
In accordance with the discussion in Sect. \ref{cuts} we see that
there is some curvature mainly generated by the $\pi\pi$ intermediate states.
In the figure the
fits linear in $s_\pi$ to the model curves below $s_\pi=0.16~$GeV$^2$
are also shown.
As can be seen the linear approximation in the relevant region
is sufficient within the precision of the present
experiment \cite{rosselet}. It becomes somewhat
borderline for the next generation of experiments but the situation improves
somewhat if we fit $|F|$ instead of the real part of $F$.
The effect of the latter we have checked in the unitarized case
where the curvature is more pronounced.

Let us now turn to the $s_\ell$ dependence.
\begin{figure}
\begin{center}
\epsfig{file=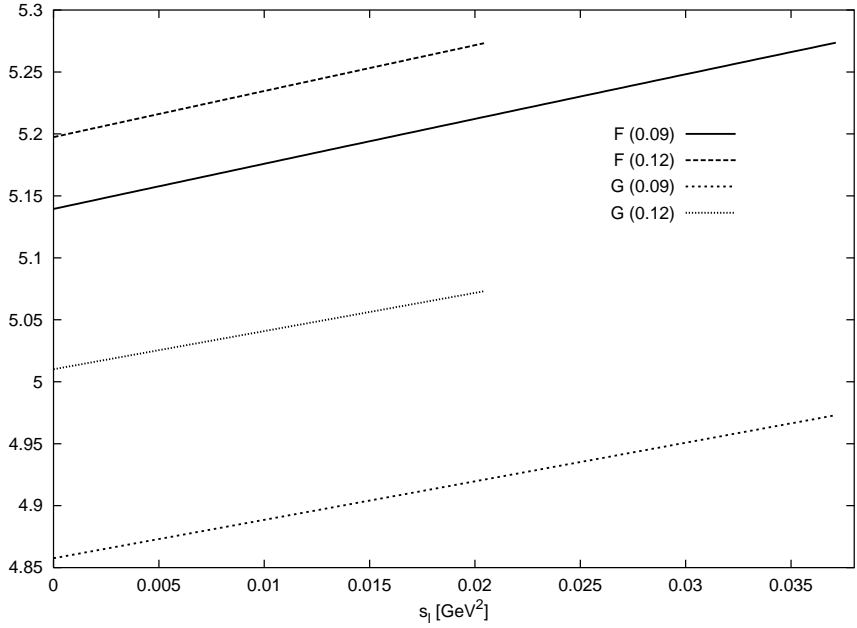,width=12cm}
\end{center}
\caption{\label{figchpt2}
The form factor $F$ and $G$ in $p^4$ CHPT
as a function of $s_\ell$ at $\cos\theta_\pi=0$ and
for $\sqrt{s_\pi}= 0.3~$GeV and $\sqrt{s_\pi}= 0.35~$GeV.}
\end{figure}
In Fig. \ref{figchpt2} the form factors $F$ and $G$ are plotted
for two values of $s_\pi$.
The lines are only plotted for the possible
kinematical domains accessible in $K_{e4}$. As is obvious from the
figure the $s_\ell$ dependence in this model is again very linear and
any curvature can be neglected
within the expected experimental accuracy.
The differences in the slope for the two values of $s_\pi$ is
small.

We now turn to the $\cos\theta_\pi$ dependence within $p^4$ CHPT.
\begin{figure}
\begin{center}
\epsfig{file=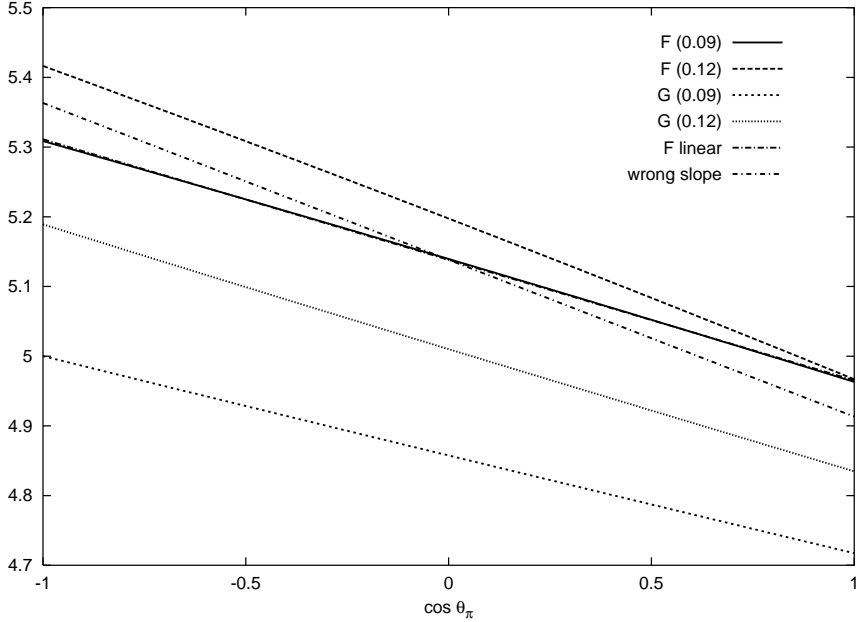,width=12cm}
\end{center}
\caption{\label{figchpt3}
The form factor $F$ and $G$ in $p^4$ CHPT
as a function of  $\cos\theta_\pi$ at $s_\ell=0$ and
for $\sqrt{s_\pi}= 0.3~$GeV and $\sqrt{s_\pi}= 0.35~$GeV.}
\end{figure}
In Fig. \ref{figchpt3} the form factors $F$ and $G$ are plotted
for two values of $s_\pi$. For $G$ it is extremely linear while for $F$
a very small curvature is present. The latter is however below
the expected experimental uncertainties and show that the effect of
$D$-waves is very small. A linear best fit to $F$ is also
shown. The slope changes somewhat between the two values of $\sqrt{s_\pi}$.
The linear best fits are :
\begin{eqnarray}
\label{fpCHPT}
F(\sqrt{s_\pi}=0.3~\mbox{GeV}) &=& 5.138-0.173 \cos\theta_\pi\nonumber\\
F(\sqrt{s_\pi}=0.35~\mbox{GeV}) &=& 5.195-0.225 \cos\theta_\pi\nonumber\\
G(\sqrt{s_\pi}=0.3~\mbox{GeV}) &=& 4.858-0.141 \cos\theta_\pi\nonumber\\
G(\sqrt{s_\pi}=0.35~\mbox{GeV}) &=& 5.011-0.177 \cos\theta_\pi
\end{eqnarray}
The differences in the $\cos\theta_\pi$ part at the
different energies are visible
but do probably remain within the errors of the next generation of
experiments.

At all relevant values of $s_\pi$ the form factors are
extremely linear in $s_\ell$ and $\cos\theta_\pi$.
Parametrizing $F$ as
$F = f_s + f_p\cos\theta_\pi$ we get from one-loop
CHPT $|f_p/f_s| = 0.043,~0.034,~0.008$
at $\sqrt{s_\pi} = 350,~300,~280$~MeV. This variation does probably stay
within the future experimental errors. 
The curvature in $s_\pi$ might become relevant in future experiments.

\section{Double Logs}
\label{doublelog}

\begin{figure}
\begin{center}
\epsfig{file=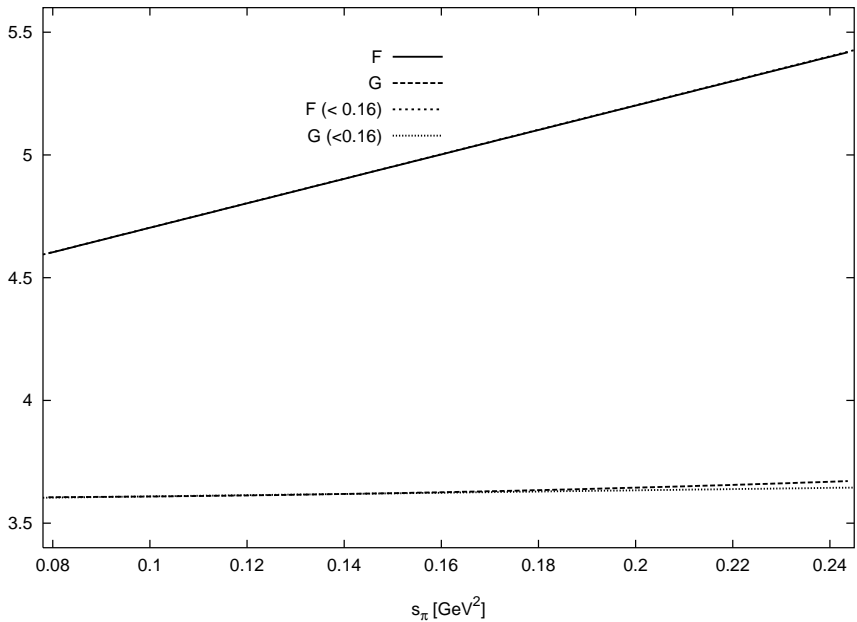,width=12cm}
\end{center}
\caption{\label{figp61}
The form factor $F$ and $G$ in partial $p^6$ CHPT
as a function of $s_\pi$ at $\cos\theta_\pi=s_\ell=0$.}
\end{figure}

\begin{figure}
\begin{center}
\epsfig{file=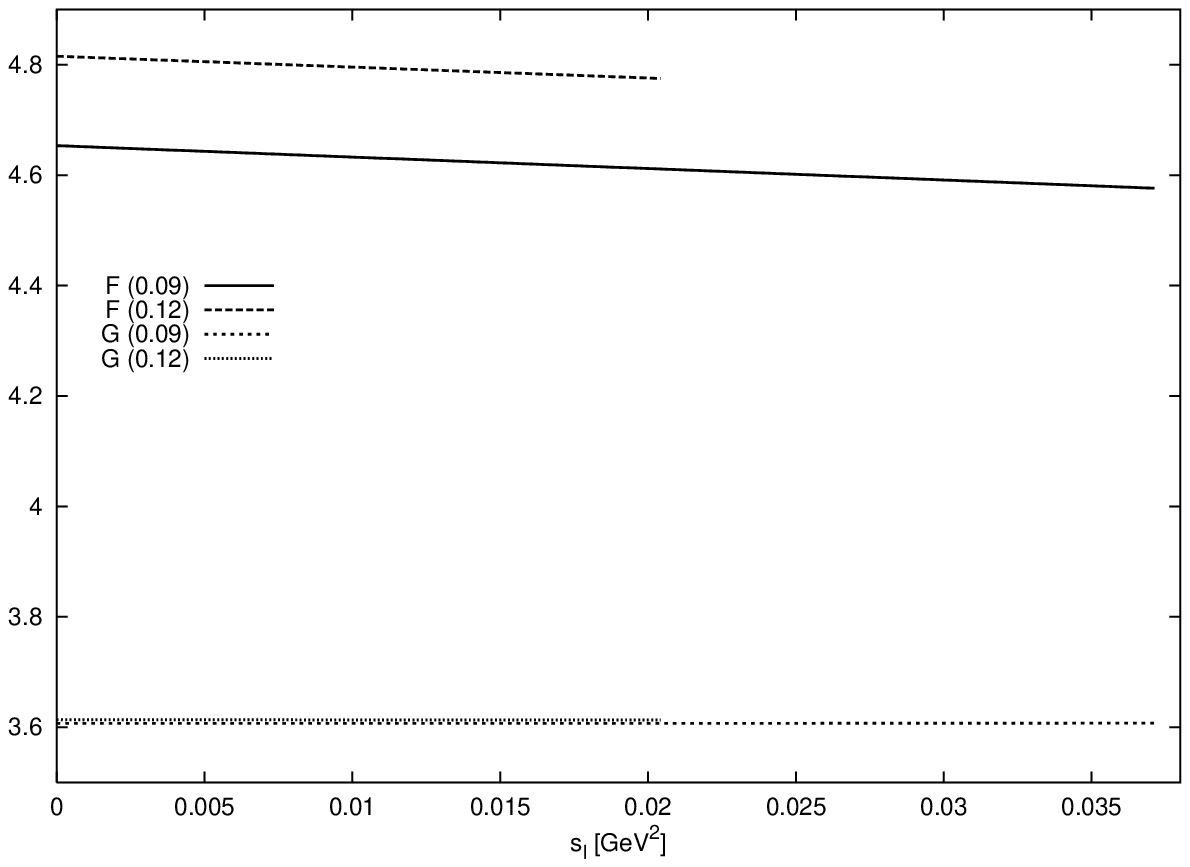,width=12cm}
\end{center}
\caption{\label{figp62}
The form factor $F$ and $G$ in partial $p^6$ CHPT
as a function of $s_\ell$ at $\cos\theta_\pi=0$ and
for $\sqrt{s_\pi}= 0.3~$GeV and $\sqrt{s_\pi}= 0.35~$GeV.}
\end{figure}

\begin{figure}
\begin{center}
\epsfig{file=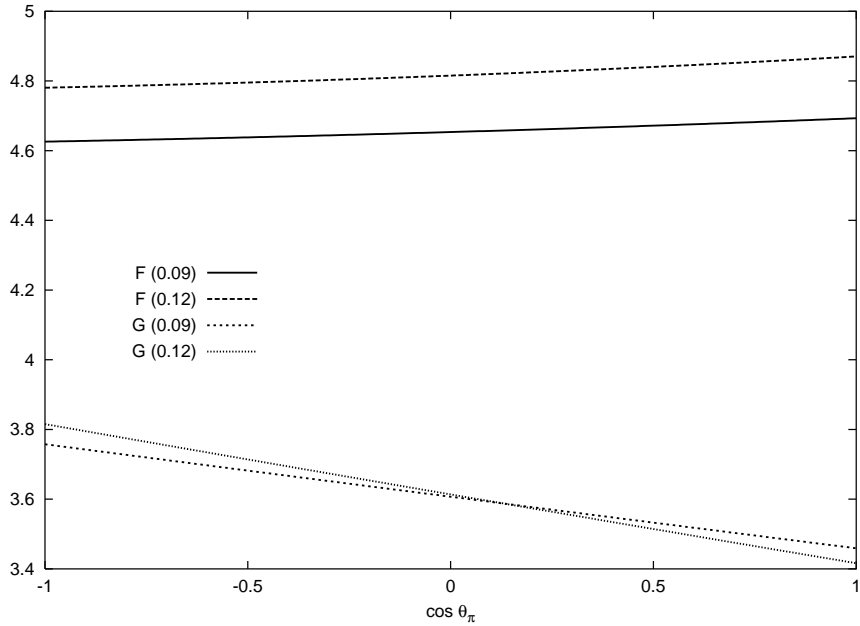,width=12cm}
\end{center}
\caption{\label{figp63}
The form factor $F$ and $G$ in partial $p^6$ CHPT
as a function of  $\cos\theta_\pi$ at $s_\ell=0$ and
for $\sqrt{s_\pi}= 0.3~$GeV and $\sqrt{s_\pi}= 0.35~$GeV.}
\end{figure}

In this section we use the partial $p^6$ calculation of \cite{BCE}
for the $F$ and $G$ form factors. Here three types of contribution have been
calculated using general renormalization methods. 
The relevant contributions are the ones proportional to
$(\log(m_K^2/\mu^2))^2$, $L_i \log(m_K/\mu)$ and $L_i L_j$.
The scale $\mu$ is the subtraction scale in CHPT and we have chosen the scale
of the logarithms to be the Kaon mass. The reason for this is that
the same approximation in the $p^4$ calculation gives a reasonable
agreement with the full result for this choice.
Figures \ref{figp61} to \ref{figp63} show the same plots as in the
previous sections. We have plotted here the partial $p^6$ results
with the lowest order added, the $p^4$ contribution is not included.
The conclusions are basically the same as in the previous sections but
notice that the curvature in $F$ and $G$ from this source is rather small.
The contribution to the linear dependence on $s_\pi$, $s_\ell$ and
$\cos\theta_\pi$ is typically smaller than the $p^4$ results
but not negligible. Absolute numerical results from this section are
quite dependent on the particular choices of input\cite{BCE}.

\section{Unitarization of $F$}

As we have seen in the plots, the behaviour of the form factors are almost 
linear in the models. Also, as we expected, the contribution from the 
$S$-wave from the $\pi\pi$ scattering produces some curvature in the 
$F$ form factor that becames more clear in the case of the one-loop
ChPT result. 
We will now discuss the $S$-wave part of 
this form factor, which is the source of the curvature
using the 
unitarization method \cite{DGL,MS}. It consists in writing a dispersive 
relation for the 
partial wave using the ChPT limit to fix the subtraction constants.
A previous  
study of the analytic properties separating the regions $s_\pi < 0$ and 
$s_\pi > 0$ provided a good way to introduce a general parametrization for 
the $S$-wave \cite{Schenk}, and the contribution from resonances 
in the t-channel (see 
\cite{BCG} for details).

\begin{figure}
\begin{center}
\epsfig{file=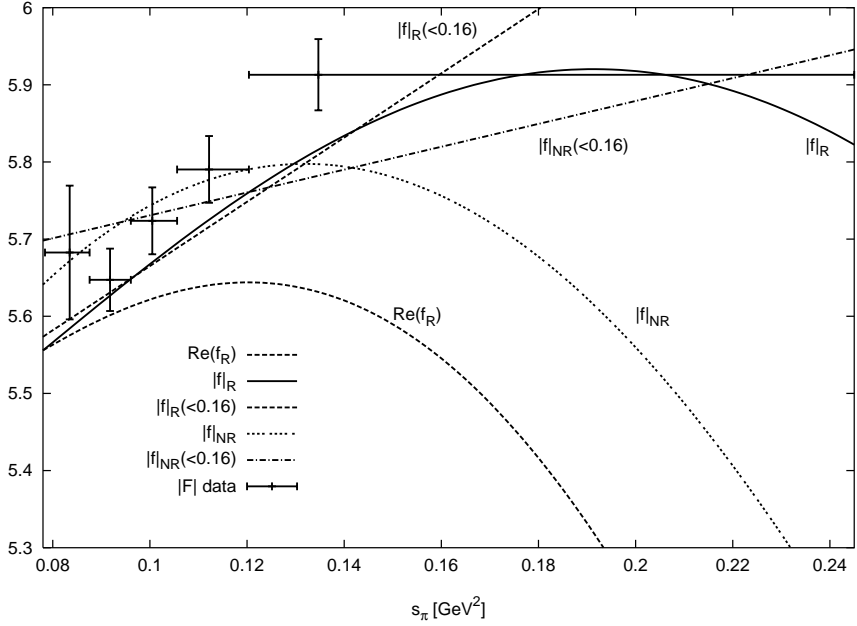,width=12cm}
\end{center}
\caption{\label{figbcg}
The S-wave in F using unitarization with ($f_R$) and without ($f_{NR}$) 
resonances as a function of $s_\pi$ at $s_\ell=0$. 
The linearized fits for $\sqrt{s_\pi}< 0.16~$GeV are also plotted.}
\end{figure}

The result is shown in Fig. \ref{figbcg}. The curve $|f|_{NR}$ represents 
the absolute value for the $S$-wave, where the contribution from 
the resonance sector, and therefore from the $s_\pi < 0$ region, is null. 
The other curves are the real and absolute value for the case with resonances.
We have shown here both the absolute value and the real part for one case
to show that  the absolute value is more linear. The latter is the relevant
one for fitting to the experimental data.

A linear fit to the relevant energies for $s_\pi$ is also plotted.
The importance of the rescattering effects in the $S$-wave is obvious, not only 
because of the modification of the linear behaviour, that can be explained by 
the fact that the cut is in the physical region, but also because of 
the contribution from the imaginary part.

\section{The $H$ form factor}
The previous sections were devoted to the form factors F and G related to the 
cuts (b), (c) and (d) of the Fig. \ref{figcuts}. It remains to discuss the 
behaviour related to the H form factor.
Using the 1-loop ChPT result, which is here order $p^6$, 
and the hypothesis of resonance saturation for the unknown $O(p^6)$ constants, 
the behaviour of this form factor is not only linear but constant around the 
experimental value $H(s_\pi=0)=-2.68 \pm 0.68$ \cite{rosselet}. There is 
a cancellation among the different contributions such that in the final 
curve the dependence with the energy is not appreciable, and the correction 
to the leading order is small \cite{Amettler}. 
Looking at the dependence with $s_\ell$ and $\cos \theta_\pi$ the shape is 
basically not modified with respect to the constant value.

\section{A sufficient parametrization}
\label{parametrization}

In all the models/approximations analyzed in the previous sections
we observed that $F$ and $G$ were always very linear in $s_\ell$ with a slope
that was very similar for both values of $s_\pi$. We also observed that 
$F$ and $G$ were very linear in $\cos\theta_\pi$ but with slopes somewhat
varying between the two energies. The form factors depend
on $s_\pi$, $t$ and $u$ and the effects of the singularities other than the
$\pi\pi$ ones can be expanded in $t$ and $u$. By expanding to first order
in these quantities we then indeed observe that both $F$ and $G$
have the structure
\begin{equation}
W = w(s_\pi) + w_1 s_\ell + w_2 \sigma_\pi X \cos\theta_\pi\,.
\end{equation}
Where we used $t-u = - 2 \; \sigma_\pi \; X \; \cos\theta_\pi$ with 
\begin{equation}
\sigma_\pi = \sqrt{1 - \frac{4 m^2_\pi}{s_\pi}}\,,\quad
X = \frac{1}{2} \lambda^{1/2}(m^2_K,s_\pi,s_\ell)\,,\quad
\lambda(a,b,c) = (a-b-c)^2-4bc\,.
\end{equation}

Checking Eqs. (\ref{fpFM}),(\ref{fpBCG}) and (\ref{fpCHPT})
we see that the $s_\pi$ dependence of the coefficient of $\cos\theta$
is indeed well described by $\sigma_\pi X$.

Using the partial wave expansions and their respective
phases $\delta^I_\ell$ (I:  isospin, $\ell$: angular momentum), 
we then obtain as a parametrization \footnote{$\tilde{f}_p$ is a combination of the 
standard $f_p$ and $g_p$ partial waves}
\begin{eqnarray}
F&=&(f_s(s_\pi)+f_1 \, s_\ell)e^{i\delta^0_0(s_\pi)}
    +\tilde{f}_p \, \sigma_\pi X\cos\theta_\pi e^{i\delta^1_1(s_\pi)}\nonumber\\
G&=&(g_p(s_\pi)+g_1 \, s_\ell)e^{i\delta^1_1(s_\pi)}
    +g_d \, \sigma_\pi X\cos\theta_\pi e^{i\delta^0_2}
\end{eqnarray}     
Then the curvature in $s_\pi$ in $G$ is mainly due to the $\rho$ resonance
and in the experimentally relevant region is rather well
described by a linear function in $s_\pi$. We therefore advocate
the use of
\begin{equation}
G = \left(g_p + g_p^\prime \, s_\pi + g_\ell \, s_\ell
\right) e^{i\delta^1_1(s_\pi)} + 
g_d \, \sigma_\pi X \cos\theta_\pi e^{i\delta^0_2(s_\pi)}
\end{equation}
The phase $\delta_1^1$
is relatively small and the last term in $G$ is already a small correction
so one can neglect the difference between $\delta_1^1$ and $\delta^0_2$
as well.
For $f_s(s_\pi)$ a linear approximation will be somewhat borderline,
inclusion of a quadratic term is therefore useful and its presence
should be checked in the data:
\begin{equation}
f_s(s_\pi) = f_s+f_s^\prime s_\pi+f_s^{\prime\prime} s_\pi^2\,.
\end{equation}
For $H$ a linear approximation in $s_\pi$ should be sufficient
\begin{equation}
H =  \left(h_p+h_p^\prime \, s_\pi\right) e^{i\delta^1_1(s_\pi)} \,.
\end{equation}
So within this approximation we have 11 parameters plus
the number of bins in $s_\pi$ for $(\delta_0^0-\delta_1^1)(s_\pi)$
as fitting parameters. 

\section{Conclusions}

We presented the results from several models for the form factors F and G
and of CHPT calculations and extensions.
We have shown that a simple linear parametrization of the dependence
on the kinematical variables is sufficient in most of the region
where there will be sufficient data with the possible exception of
the $S$-wave part of the $F$ form-factor. Even there a linear fit
is quite a good approximation but then care has to be taken to
define the theoretical slope.

We showed that 11 parameters in addition to the phases are quite sufficient
to fit the expected accuracy of the experiment and this should allow
to determine those phases more accurately as a function of $s_\pi$ compared
to the case where everything is fitted on a bin-by-bin basis.

\section{Acknowledgements}
We thank  Juerg Gasser, Bachir Moussallam and Stefan Pislak
for useful comments.


\begin{thebibliography}{99}
\bibitem{rosselet} L. Rosselet et al., Phys. Rev. D15 (1977) 574.
\bibitem{kl4daphne} J. Bijnens et al., hep-ph/9411311, chapter 7.1
in ``The Second
DA$\Phi$NE PHYSICS HANDBOOK'',
eds. L. Maiani, G. Pancheri and N. Paver, INFN, Frascati.
\bibitem{BCG} J.~Bijnens, G.~Colangelo and J.~Gasser, 
  Nucl. Phys. B427 (1994) 427 (hep-ph/9403390).
\bibitem{chptp4} J. Bijnens, Nucl. Phys. B337 (1990) 635;\\
C.~Riggenbach et al., Phys. Rev. D43 (1991) 127.
\bibitem{Chounet} L.-M. Chounet, J.-M. Gaillard and M.K. Gaillard,
Phys. Rep. C4 (1972) 199.
\bibitem{Pais} A.~Pais and S.B.~Treiman, Phys. Rev. 168 (1968) 1858.
\bibitem{PaisTreiman} G. Colangelo, M. Knecht and J. Stern, Phys.Lett.
B336 (1994) 543 (hep-ph/9406211). 
\bibitem{Amettler} Ll. Amettler et al., Phys. Lett. B303 (1993) 140.
\bibitem{EGPR} G.~Ecker et al., Nucl. Phys. B321 (1989) 311.
\bibitem{FM} M. Finkemeier and E. Mirkes,
  Z. Phys. C69 (1996) 243 (hep-ph/9503474).
\bibitem{introdaphne} J.~Bijnens, G.~Ecker and J.~Gasser, hep-ph/9411232,
chapter 3 in ``The Second
DA$\Phi$NE PHYSICS HANDBOOK'',
eds. L. Maiani, G. Pancheri and N. Paver, INFN, Frascati.
\bibitem{BCE} J.~Bijnens, G.~Colangelo and G.~Ecker, 
Phys. lett. B441 (1998) 437 (hep-ph/9808421).
\bibitem{DGL} J.~Donoghue, J.~Gasser and H.~Leutwyler, Nucl. Phys. B343 
(1990) 341.
\bibitem{MS} A.D.~Martin and T.D.~Spearman, Elementary particle theory 
(North-Holland, Amsterdam, 1970). 
\bibitem{Schenk} A.~Schenk, Nucl. Phys. B363 (1991) 97.
\end{thebibliography}
\end{document}